\pdfoutput=1
\documentclass[twocolumn,superscriptaddress,nofootinbib]{revtex4-1}
\usepackage{graphics, graphicx, amsmath, amssymb}
\usepackage{epstopdf}
\usepackage{dsfont}
\usepackage{hyperref}
\usepackage{color}
\usepackage[table,xcdraw]{xcolor}
\usepackage[normalem]{ulem}
\newcommand{\eq}[1]{\begin{equation}\begin{aligned}#1\end{aligned}\end{equation}}
\newcommand{\iu}{\text{i}}
\newcommand{\eu}{\text{e}}
\newcommand{\ha}{\hat{a}}
\newcommand{\had}{\hat{a}^\dagger}
\newcommand{\hb}{\hat{b}}

\newcommand{\ket}[1]{\left|#1\right\rangle}
\newcommand{\bra}[1]{\left\langle#1\right|}
\newcommand{\braket}[2]{\left.\left\langle#1\right|#2\right\rangle}

\newcommand{\add}[1]{{#1}}
\newcommand{\rem}[1]{}

\begin{document}

\title{Transcoherent states:  Optical states for maximal generation of atomic coherence}
\author{Aaron Z. Goldberg}
\email{goldberg@physics.utoronto.ca}
\affiliation{Department of Physics and Centre for Quantum Information \& Quantum Control, University of Toronto, Toronto, Ontario, Canada M5S 1A7}
\author{Aephraim M. Steinberg}
\affiliation{Department of Physics and Centre for Quantum Information \& Quantum Control, University of Toronto, Toronto, Ontario, Canada M5S 1A7}
\affiliation{CIFAR, 661 University Ave., Toronto, Ontario M5G 1M1, Canada}

\begin{abstract}
    {Quantum technologies are built on the power of coherent superposition}. Atomic 
    {coherence is} typically generated from optical coherence, most often via Rabi oscillations. However, canonical coherent states of light create imperfect resources; a fully-quantized description of ``$\tfrac{\pi}{2}$ pulses'' shows that the atomic superpositions generated remain entangled with the light.
    We show that there are quantum states of light that 
    generate coherent atomic states {perfectly,} with no residual atom-field entanglement. These states can be found for arbitrarily short times and approach slightly-number-squeezed $\tfrac{\pi}{2}$ pulses in the limit of large intensities{; similar ideal states can be found for any $(2k+1)\tfrac{\pi}{2}$ pulses, requiring more number squeezing with increasing $k$.} 
    Moreover, these states can be repeatedly used as ``quantum catalysts''
    to successfully generate coherent atomic states with high probability.
    From this perspective we have identified states that are {``}more coherent{"} than coherent states.
\end{abstract}

\maketitle

Coherence is {the quintessential} property of quantum systems. Underlying interference, coherence plays a role in optics \cite{Sudarshan1963,Glauber1963}, atomic physics \cite{Kasevich2002,Bloch2008}, and beyond{, and in multipartite systems is also the defining feature of entanglement}. Our exquisite ability to manipulate coherence paves the way for applications from quantum information to quantum thermodynamics, and likewise the success of these applications depends on our ability to prepare and measure quantum coherence. 
{In this paper, w}e 
consider the limits of standard approaches
for transferring maximal coherence from light to atoms and identify optimal states of light with coherence properties that outperform all standard approaches.

That quantum mechanical systems permit coherent superpositions provides a resource for quantum tasks \cite{LostaglioJenningsRudolph2015,LostaglioKorzekwaetal2015}. For example, quantum coherence can be used to drive energy transfer between two systems at thermal equilibrium \cite{Korzekwaetal2016,Messingeretal2020}, a task forbidden by classical thermodynamics, helping launch the nascent field of quantum thermodynamics \cite{GemmerMahler2009,kosloff2013,Brandaoetal2015,VinjanampathyAnders2016}.
This realm of ideas led to significant development of resource theories for quantum coherence, which quantify the usefulness of quantum states for achieving various tasks \cite{Aberg2006,LeviMintert2014,Baumgratzetal2014,WinterYang2016,Zhangetal2016,Streltsovetal2017}. 

In the simplest quantum system, a two-level atom described by ground and excited states $\ket{g}$ and $\ket{e}$, the most useful states are pure-state superpositions with equal probability of being found in either energy level. In some ways, these states behave like $50:50$ probabilistic mixtures of ground and excited states, while the states' coherence properties confer additional wavelike behaviour that cannot be described by probabilistic mixtures (see Ref. \cite{Bartlettetal2006} for an interpretational discussion of this distinction). Such superpositions and their corresponding large dipole moments enable, among many other examples, the aforementioned quantum engines \cite{Korzekwaetal2016},
quantum thermometers \cite{Jevticetal2015,Thametal2016,Mancinoetal2017,Maureretal2019}, \add{entanglement generation \cite{Hensenetal2015},}
teleportation and quantum cryptographic schemes \cite{LeeKim2000,Leeetal2003}, quantum logic gates using Rydberg atoms \cite{Mulleretal2009,ParedesBaratoAdams2014,Mulleretal2014,Malleretal2015,Levineetal2018}, and even the creation of superpositions of atoms and molecules \cite{Dowlingetal2006}.
The interactions between these atoms and light can be controlled exceedingly well in cavities \cite{Raimondetal2001}, including being implemented in cavities made from quantum dots \cite{Yoshieetal2004} and superconducting circuits \cite{Finketal2008,Blaisetal2020}.

The typical method for producing this resource is an interaction between a ``coherent state'' of light \cite{Sudarshan1963,Glauber1963} and an atom in its ground state. 
{When an atom interacts with a classical field near resonance, it behaves like a spin in the presence of a magnetic field, exhibiting ``Rabi flopping'' \cite{AllenEberly1987}.} 
{This leads to coherent oscillations} between $\ket{g}$ and $\ket{e}$ at the Rabi frequency $\Omega_0 \sqrt{\bar{n}}$, where $\bar{n}$ is the intensity of the field in units of single-photon intensity, such that a ``$\tfrac{\pi}{2}$ pulse'' perfectly transfers the atom into a coherent superposition of $\ket{g}$ and $\ket{e}$ after a time $\Omega_0\sqrt{\bar{n}} t=\tfrac{\pi}{2}$.  
According to this semiclassical description, {coherence between $\ket{g}$ and $\ket{e}$ is perfect, and the oscillations continue indefinitely in the absence of spontaneous emission or other broadening.
But this picture is modified by quantization of the electromagnetic field, the quantum uncertainty in photon number leading to a spread in Rabi frequencies, and thus a progressive decay of the oscillations; the {\it discrete} nature of these Rabi frequencies is at the origin of the famous ``collapses and revivals'' of Rabi oscillations~\cite{Eberlyetal1980}, one of the phenomena long studied for its role in experimentally demonstrating the quantum nature of light~\cite{Rempeetal1987}.}

{The quantum analog of a $\tfrac{\pi}{2}$ pulse is intrinsically imperfect.  Since the pulse area is proportional to $\sqrt{\bar{n}}$, an uncertainty in photon number amounts to an uncertain pulse area, and atomic coherence that never attains its theoretical maximum; on the other hand, although the ${\Delta}n\rightarrow0$ limit leads to perfect coherence between $\ket{{g,n}}$ and $\ket{e,{n-1}}$, the fact that the field has complete {\it welcher Weg} information about the state of the atom means that the atom is in a completely classical, incoherent, mixture of $\ket{g}$ and $\ket{e}$.  In both limits, entanglement between the atomic state and the state of the light prevents the atom from being completely coherent.
{(For pioneering studies of the atom-field entanglement in Jaynes-Cummings settings, see for instance \cite{GeaBanacloche1990,PhoenixKnight1991a,PhoenixKnight1991b,GeaBanacloche1991,GeaBanacloche2002,vanEnkKimble2002,SilberfarbDeutsch})
}.}

{Below, we address the question of what the maximum coherence transferable from light to an atom is, and which quantum states of the light achieve this.}
We find {quantum states that can produce \textit{perfect} atomic coherence with no residual atom-field entanglement},
but that these states  are markedly different from the canonical coherent states -- although they share some important properties. 
The \textit{transcoherent states} in this optimal family generalize $\tfrac{\pi}{2}$ pulses to quantum descriptions of light and can be used as precursors, or maybe even catalysts \cite{Aberg2014,Messingeretal2020},\footnote{{The catalysts of quantum catalysis are changed by backaction and might be termed pseudo-catalysts.}} for creating optimally-coherent atomic states.

\section{Theoretical context}
We are seeking states with maximal coherence (off-diagonal matrix elements) in the energy eigenbasis; i.e., states that maximize the absolute value of the dipole moment $\ket{e}\bra{g}$.  
Maximally coherent states have the form \eq{\frac{\ket{g}+\eu^{\iu\phi}\ket{e}}{\sqrt{2}},
\label{eq:atomic coherent state}
} {where $\phi$ is any real phase difference}. Free evolution of the atomic system leads to a periodic evolution in $\phi$ at the resonance frequency $\omega$, so creating an atomic state with \textit{any} $\phi$ is sufficient for creating an atomic state with arbitrary $\phi$.

{We consider a single mode of the electromagnetic field with zero detuning from the atomic resonance; its coupling to the atom can be treated using the well-known Jaynes-Cummings Hamiltonian:}
\eq{H=\add{\omega\left(\had\ha +\ket{e}\bra{e}\right)+}\frac{\Omega_0}{2}\left(\ha\sigma_-+\had\sigma_+\right),} where $\Omega_0$ is the vacuum Rabi frequency, $\sigma_\pm$ is the atomic raising (lowering) operator $\ket{e}\bra{g}$ ($\ket{g}\bra{e}$), and $\ha$ ($\had$) annihilates (creates) an excitation in the field mode. This interaction preserves total excitation number, coupling pairs of atom-field states $\ket{g}\otimes\ket{n+1}$ and $\ket{e}\otimes\ket{n}$, where $\ket{n}$ denotes a field state with exactly $n$ photons. A general atom-field state as a function of time is given by \eq{\ket{\psi\left(t\right)}=c_{g,0}\ket{g}\otimes\ket{0}+\sum_{n=0}^{\infty}c_{g,n+1}\left(t\right)\ket{g}\otimes\ket{n+1}
\\
+c_{e,n}\left(t\right)\ket{e}\otimes\ket{n}.} From this general expression we can evaluate the atomic coherence at any time for any set of initial conditions.

It is easy to see that if one begins with the field state that \textit{would be} produced by \eqref{eq:atomic coherent state} interacting with the vacuum for a single-excitation Rabi period $\Omega_0t=\pi$ (a ``{single-excitation} $\pi$ pulse''),
\eq{
\frac{\ket{0}+\eu^{\iu\phi}\ket{1}}{\sqrt{2}},
\label{eq:field 0 plus 1}
} one can wait the duration of a {single-excitation} $\pi$ pulse to perfectly create an coherent atomic state. The coherence present in the optical state, together with linearity, leads to perfect coherence transfer to the atom.  {Of course, the single-photon Rabi frequency is generally quite low, meaning that it would take a long time to generate coherence in this way.  To transfer coherence more rapidly, higher Rabi frequencies, and thus higher photon numbers, are required.} We presently show how this procedure generalizes to optical states with more than one photon.

There are many equivalent ways of expressing the coherence in a two-level atom, most of which involve the atomic reduced density matrix
\eq{
\rho(t)=\sum_n \braket{n}{\psi(t)}\braket{\psi(t)}{n}
.} If the reduced state is pure and equal to \eqref{eq:atomic coherent state} for some $\phi$, which without loss of generality can be absorbed into the definition of $\ket{e}$, we say that the atomic state is perfectly coherent.  Thus, a simple-to-state goal for creating a coherent atomic state is that $\rho(t)$ be the $+1$ eigenstate of a projection operator formed from \eqref{eq:atomic coherent state}, and we explicitly set $\phi$ to 0. A measure of coherence could thus be the probability of measuring the system to be in one of the eigenstates:
\eq{
P(t)=\frac{\bra{g}+\bra{e}}{\sqrt{2}}\rho(t) \frac{\ket{g}+ \ket{e}}{\sqrt{2}}.
\label{eq:probability measure}
} Another measure of coherence is given by the size of the off-diagonal elements of $\rho(t)$ \cite{Baumgratzetal2014}:
\eq{
\mathcal{C}(t)=\left|\bra{e}\rho(t)\ket{g}\right|+\left|\bra{g}\rho(t)\ket{e}\right|
&=2\left|\sum_{n=0}^\infty c_{e,n}^*\left(t\right)c_{g,n}\left(t\right)\right|
\\&=2 P(t)-1
.
\label{eq:coherence measure}
} The properties of $2\times 2$ density matrices enforce the bounds $0\leq \mathcal{C},P \leq 1$, allowing these metrics to evaluate the coherence transferred by non-optimal and near-optimal states of light.

The eigenstates of $H$ are simple to state:
\eq{
\ket{\pm,n}=\frac{1}{\sqrt{2}}\ket{e}\otimes\ket{n}\pm\frac{1}{\sqrt{2}}\ket{g}\otimes\ket{n+1}
.} 
Combinations of the eigenstates $\ket{\pm,n}$ transfer probability back and forth at quantized Rabi frequencies, which increase with $n$:
\eq{\Omega_n=\Omega_0\sqrt{n+1};} 
states with different total excitation numbers $n+1$ are not coupled by the JCM.
An initially-separable atom-field state with exactly $n+1$ excitations is again separable after the two subsystems completely swap an excitation, which happens over the duration of a quantized $\pi$ pulse $\Omega_n t=\pi$. 
These oscillations in probability are exactly the same as in the semiclassical treatment of the problem 
{(when $\bar{n}$ is taken to be the total excitation number, i.e., the photon number when the atom is in the ground state).}
However, for the duration of the Rabi period, the atom and field are entangled, and the atomic coherence vanishes for all time $\mathcal{C}(t)=0$. Superpositions of states with different numbers of excitations are necessary in order to build up atomic coherence. 

\section{Creating atoms with maximal coherence}

The most experimentally-friendly state from which to create maximal atomic coherence is an atom in its ground state. 
The simplest method for creating maximal coherence couples an atom in its ground state with the superposition state given in \eqref{eq:field 0 plus 1}. After half of a vacuum Rabi period, the excitation in $\ket{g}\otimes\ket{1}$ is entirely converted into an atomic excitation $\ket{g}\otimes\ket{0}$, while the $\ket{g}\otimes\ket{0}$ component has no excitations to transfer. A direct generalization beyond this known state fails: the initial state
\eq{
\ket{g}\otimes\frac{\ket{n}+\ket{n+1}}{\sqrt{2}}
} only generates coherence \eq{
\mathcal{C}(t)=\left|\sin\left(\frac{\Omega_n t}{2}\right)\cos\left(\frac{\Omega_{n-1} t}{2}\right)\right|,
} which equals 1 if and only if $n=0$ and $\sin\left(\frac{\Omega_0 t}{2}\right)=1$. The presence of different quantized Rabi frequencies $\Omega_n$ whose ratio is irrational prohibits perfect coherence generation.

{Yet it is straightforward to generalize this idea and determine which field states create the most coherence, by inspecting the general solutions to the JCM beginning in the ground state}: 
\eq{
\ket{g}\otimes \sum_{n=0}^\infty c_n \ket{n} \to
\ket{\psi(t)}= c_0\ket{g}\otimes\ket{0}+\sum_{n=0}^{\infty}c_{n+1}\eu^{-\iu n\omega t}
\\\times
\left[	
\cos\left(\frac{\Omega_n t}{2}\right)\ket{g}\otimes\ket{n+1}-\iu \sin\left(\frac{\Omega_n t}{2}\right)\ket{e}\otimes\ket{n}
\right].
\label{eq:JCM solution from ground}
\color{black}
} 
The coherence measure \eqref{eq:coherence measure} is given by
\eq{\mathcal{C}\left(t\right)&=
2\left|\sum_{n=0}^\infty c_{n+1}^*c_{n}\sin\left(\frac{\Omega_n t}{2}\right)\cos\left(\frac{\Omega_{n-1} t}{2}\right)\right|.
\label{eq:coherence measure solution}
}
An ideal field state must balance {on the one hand} a narrow distribution of Rabi frequencies {in order to keep} $\sin\left(\frac{\Omega_n t}{2}\right)\cos\left(\frac{\Omega_{n-1} t}{2}\right)\approx \tfrac{1}{2}$ {-- corresponding to a narrow distribution $\left|c_n\right|^2$ -- and on the other hand} a large overlap $\left|c_{n+1}c_n\right|$ -- corresponding to a broad distribution $\left|c_n\right|^2$. Ideally, the peak of the distribution will occur at the maximum of the trigonometric terms, and the trigonometric terms will vary slowly near this point. In the large-$\bar{n}$ limit, this is achieved by $\tfrac{\pi}{2}$ pulses with $\Omega_0 t \sqrt{\bar{n}}\approx \tfrac{\pi}{2}$, just like the classical solution:
near this $\bar{n}$,
the trigonometric terms can be approximated by 
\eq{
\sin\left(\frac{\Omega_{\bar{n}}  {t}}{2}\right)\cos\left(\frac{\Omega_{\bar{n}-1} {t}}{2}\right)\approx \tfrac{1}{2}\left[1 +\frac{\pi}{8\bar{n}}+\mathcal{O}\left(\frac{1}{\bar{n}^2}\right)\right] 
\label{eq:trig terms coherence}
.} 

The previous consideration sets the mean of the photon-number distribution to correspond to $\tfrac{\pi}{2}$ pulses; next we optimize its variance. If the product $c_{n+1}^*c_n$ is proportional to $1-\tfrac{\pi}{8\bar{n}}$, this matches the $\mathcal{O}\left(\frac{1}{\bar{n}}\right)$ term such that $\mathcal{C}$ deviates from unity as $\mathcal{O}\left(\frac{1}{\bar{n}^2}\right)$.
A balance can thus be found by states whose photon-number variance $\sigma^2$ achieves this match:
\eq{
\left|c_{\bar{n}+\delta}\right|^2&=|c_{\bar{n}}|^2 \exp\left(-\frac{\delta^2}{2\sigma^2}\right)\\
&\Rightarrow
 |c_{\bar{n}+1}c_{\bar{n}}|=|c_{\bar{n}}|^2\eu^{-\frac{1}{4\sigma^2}}\approx |c_{\bar{n}}|^2\left(1-\frac{1}{4\sigma^2}\right)
;
\label{eq:number squeezed Gaussian}
} the $\mathcal{O}\left(\tfrac{1}{\bar{n}}\right)$ term of $\mathcal{C}$ vanishes when the variance is $\sigma^2=\tfrac{2\bar{n}}{\pi}$.
This number squeezing by a factor of $\tfrac{2}{\pi}$ relative to coherent states with $\sigma^2=\bar{n}$, the optimal amount found by Ref. \cite{Messingeretal2020}, is best matched to the slightly-broadened distribution of the trigonometric coefficients \eqref{eq:trig terms coherence}.  

This number-squeezing result is approximate; {remarkably,} we can find a {\it perfect} balance by further examining \eqref{eq:JCM solution from ground}. In each photon-number subspace we compute
\eq{
&\braket{n}{\psi(t)}=\eu^{-\iu n\omega t}\\
&\quad\times
\left[c_n \eu^{\iu\omega t}\cos\left(\frac{\Omega_{n-1} t}{2}\right)\ket{g}-\iu \sin\left(\frac{\Omega_n t}{2}\right) c_{n+1}\ket{e}\right].
} The desired atomic states require a series of pairs of equalities between adjacent excitation manifolds:
\eq{
c_n \eu^{\iu\omega t}\cos\left(\frac{\Omega_{n-1} t}{2}\right)=-\iu \sin\left(\frac{\Omega_n t}{2}\right) c_{n+1}
,\label{eq:recursion}
}
which, for normalization constant $\mathcal{N}=\sum_{n=0}^\infty \left|c_n\right|^2\cos^2\left(\tfrac{\Omega_{n-1}t}{2}\right)$, leads to \eq{
\ket{\psi(t)}=\frac{\ket{g}+\ket{e}}{\sqrt{2}}\otimes \frac{1}{\sqrt{\mathcal{N}}}\sum_{n=0}^\infty c_n \eu^{-\iu n \omega t} \cos\left(\frac{\Omega_{n-1} t}{2}\right)\ket{n}.
\label{eq:ground to atomic coherent}
} Equation \eqref{eq:recursion} defines a recursion relation for the field-state coefficients. Whenever this series truncates,
an arbitrary perfectly-coherent state \eqref{eq:atomic coherent state} can be created that is completely separable from the optical field and can be used for quantum tasks.

These generalized field states achieve perfect coherence generation, and are thereby transcoherent states, whenever \eqref{eq:recursion} truncates and $\mathcal{N}$ is finite. This is equivalent to the highest-excitation manifold undergoing a $\pi$ pulse. Since the interaction time $t$ can be controlled experimentally, there is an infinite family of solutions \eq{
t_{n_\text{max}}=\frac{\pi}{\Omega_{n_\text{max}-1} }
\label{eq:quantum pi pulses}
} that together with \eqref{eq:recursion} achieve perfect coherence generation. At $t_1$ we exactly recover the earlier solution \eqref{eq:field 0 plus 1} and single-excitation $\pi$ pulse $\Omega_0 t=\pi$; our results directly generalize this by adding increasingly-many terms to the initial superposition, corresponding to quantized $\pi$ pulses for the maximal-excitation-number manifold $\Omega_0 t_{n_\text{max}}=\tfrac{\pi}{\sqrt{n_\text{max}}}$ so that there is no probability of measuring the state to be $\ket{g}\otimes\ket{n_\text{max}}$.
That the recursion relation \eqref{eq:recursion} does not diverge is equivalent to none of the intermediate excitation manifolds undergoing any integer multiple of a $\pi$ pulse. The $n$th excitation manifold must contribute to both adjacent photon-number subspaces $\ket{n}$ and $\ket{n-1}$ for all $n<n_\text{max}$, and the $n_\text{max}$th-excitation manifold must only contribute to the $(n_\text{max}-1)$-photon subspace.

\begin{figure}
    \centering
    \includegraphics[width=\columnwidth]{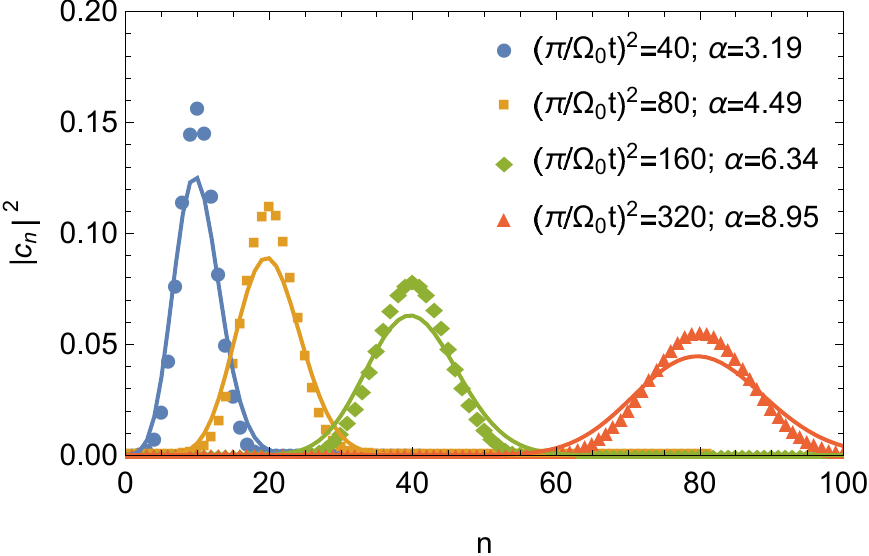}
    \caption{Probability distribution for ``transcoherent'' field states that perfectly create coherent atomic states from $\ket{g}$ in time $t$. Also plotted are the (Poissonian) probability distributions for coherent states that are classically used for $\tfrac{\pi}{2}$ pulses. The ideal field state distributions approach slightly-number-squeezed coherent states as $t$ decreases and the average energy increases.}
    \label{fig:amplitudes ideal from ground}
\end{figure}

Of note, the times for coherence transfer correspond to $\pi$ pulses for the \textit{maximal} photon numbers, while the classical optimal solutions correspond to $\tfrac{\pi}{2}$ pulses for the \textit{average} photon numbers.
The coherence is transferred in a time that scales with $\tfrac{1}{\sqrt{\bar{n}}}$, increasing in speed with increased input energy. {This is faster by a factor of $\bar{n}$ than the approximate solutions found in Refs. \cite{GeaBanacloche1990,GeaBanacloche1991,PhoenixKnight1991a,PhoenixKnight1991b}.} \add{Those works found that atoms interacting with strong coherent states eventually attain a large amount of coherence at a time halfway between the ``collapse'' and ``revival'' of Rabi oscillations. While their results were significant because coherence can be found long after it seems to have disappeared, our optimal result shows that coherence can be generated significantly more quickly, and can be achieved even without intense beams of light.} We 
plot some of these optimal states in Fig. \ref{fig:amplitudes ideal from ground}. 

\section{Other transcoherent states}
Before exploring the properties and ramifications of maximal coherence transfer we comprehensively show other methods of achieving the same goal that are less experimentally friendly. Classically, one can create the state \eqref{eq:atomic coherent state} using not only $\tfrac{\pi}{2}$ pulses, but also $(2k+1)\tfrac{\pi}{2}$ pulses for any integer $k$. It turns out that for even $k$ there exist transcoherent field states that will perfectly transfer coherence to $\ket{g}$, and for odd $k$ there exist transcoherent field states that will perfectly transfer coherence to $\ket{e}$. All of these pulses take longer than our initially-mentioned $\tfrac{\pi}{2}$ pulses, and more average energy in the field states, so we will later focus our analysis on the original set of pulses.

Every excitation manifold with states $\ket{\pm,n}$ can contribute to photon-number subspaces with adjacent values $n$ and $n+1$. To achieve perfect coherence transfer, the lowest-excitation manifold must not contribute any amplitude to the state $\ket{e}\otimes \ket{n_\text{min}-1}$, and the highest-excitation manifold must not contribute any to the state $\ket{g}\otimes \ket{n_\text{max}+1}$. Our earlier states ensured this property by achieving no evolution in the zero-excitation manifold and a $\pi$ pulse in the highest-excitation manifold.

When the atom begins in $\ket{g}$ we can achieve perfect coherence transfer when the lowest-excitation manifold undergoes a $(2k)\pi$ pulse and the highest a $(2k+1)\pi$ pulse. As long as the intervening coefficients satisfy the recursion relation \eqref{eq:recursion}, the atomic state \eqref{eq:atomic coherent state} will be perfectly created. The solutions for $k>0$ correspond to the periodic stationary points of the recursion relation \eqref{eq:recursion} and the corresponding periodic maxima of the trigonometric terms [c.f. \eqref{eq:trig terms coherence} and Appendix \ref{app:how much squeezing}].

\begin{figure}
    \centering
    \includegraphics[width=\columnwidth]{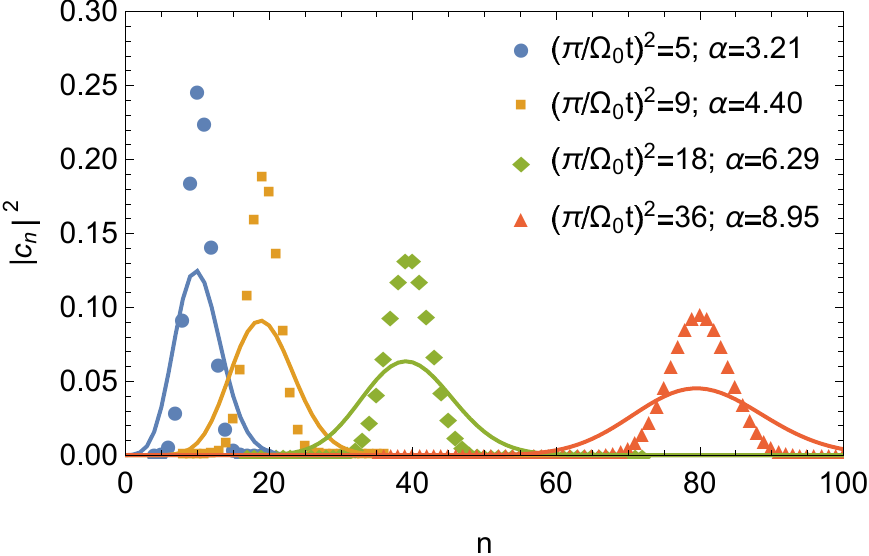}
    \caption{Probability distribution for field states that perfectly create coherent atomic states from $\ket{e}$ in time $t$. Also plotted are the (Poissonian) probability distributions for coherent states that are classically used for $\tfrac{3\pi}{2}$ pulses. The ideal field state distributions approach number-squeezed coherent states as $t$ decreases and the average energy increases; the number squeezing is much more significant than in Fig. \ref{fig:amplitudes ideal from ground}.
    }
    \label{fig:amplitudes ideal from excited}
\end{figure}

When the atom is initially excited, perfect coherence transfer can be achieved with a $(2k+1)\pi$ pulse in the lowest-excitation manifold that is simultaneously a $(2k+2)\pi$ pulse in the highest. 
The atom-field state $\ket{e}\otimes\sum_n c_n\ket{n}$ evolves to
\eq{
&\ket{\psi(t)}=
\sum_{n=0}^\infty\eu^{-\iu n\omega t}\ket{n}
\\&\quad\otimes
\left[c_n\eu^{-\iu \omega t}\cos\left(\frac{\Omega_n t}{2}\right)\ket{e}-c_{n-1}\iu\sin\left(\frac{\Omega_{n-1} t}{2}\right)\ket{g}\right]
,
} where the $c_{-1}$ coefficient can safely be ignored because $\Omega_{n-1}=0$. The recursion relation
\eq{
c_{n}=c_{n-1}\frac{-\iu\eu^{\iu\omega t}\sin\left(\frac{\Omega_{n-1} t}{2}\right)}{\cos\left(\frac{\Omega_{n} t}{2}\right)}
\label{eq:recursion starting e}
} is required to 
make $\braket{n}{\psi(t)}$ {maximally} coherent for $n_\text{min}\leq n<n_\text{max}$. This, together with a time $\Omega_{n_{\text{min}}} t=(2k+1)\pi$, perfectly produces a coherent atomic state from an initially-excited atomic state (see Fig. \ref{fig:amplitudes ideal from excited}).

The above results can also be concatenated to yield new transcoherent states that perfectly generate \eqref{eq:atomic coherent state}. 
For example, when the atom begins in $\ket{g}$ and we wait a time $\Omega_0 t\sqrt{n_\text{max}}=\pi$, the coefficients between any $c_{(2k)^2 n_\text{max}}$ and $c_{(2k+1)^2n_\text{max}}$ can be populated according to the recursion relation \eqref{eq:recursion}, so long as the coefficients between any $c_{(2k+1)^2 n_\text{max}}$ and $c_{(2k+2)^2n_\text{max}}$  are always zero.
Equivalently, when the atom begins in $\ket{e}$ and we wait a time $\Omega_0 t\sqrt{n_\text{min}+1}=\pi$, the coefficients between any $c_{(2k+1)^2(n_\text{min}+1)-1}$ and $c_{(2k+2)^2(n_\text{min}+1)-1}$ can be populated according to the recursion relation \eqref{eq:recursion starting e}, so long as the coefficients between $c_{(2k)^2(n_\text{min}+1)-1}$ and $c_{(2k+1)^2(n_\text{min}+1)-1}$ are always zero. Superpositions of these concatenated solutions comprise the full set of transcoherent field states that can be used to perfectly generate coherent atomic states.

\section{Investigating the transcoherent states that yield perfect $\tfrac{\pi}{2}$ pulses}

Perfect coherence generation occurs at a discrete set of times $\Omega_0 t_{n_\text{max}}=\pi/\sqrt{n_\text{max}}$ corresponding to $\pi$ pulses for each possible maximum photon number. Although this set approaches a continuum as $n_\text{max}$ grows, and one can always turn off the interaction at some $t_{n_\text{max}}$ and wait $t-t_{n_\text{max}}$ in order to achieve perfect atomic coherence at an arbitrary time $t$, there are other strategies for maximizing the coherence generated at arbitrary times. Crucial insight comes from our perfect solutions \eqref{eq:recursion} and the form of the coherence measure \eqref{eq:coherence measure solution}. The coherence measure looks like an inner product between vectors with components $c_{n}\cos\left(\frac{\Omega_{n-1} t}{2}\right)$ and $c_{n+1}\sin\left(\frac{\Omega_n t}{2}\right)$; the Cauchy-Schwarz inequality dictates that the recursion relation \eqref{eq:recursion} maximizes this inner product. This realization can be applied at any time $t$ so long as the vectors are normalized.

\begin{figure}
    \centering
    \includegraphics[width=\columnwidth]{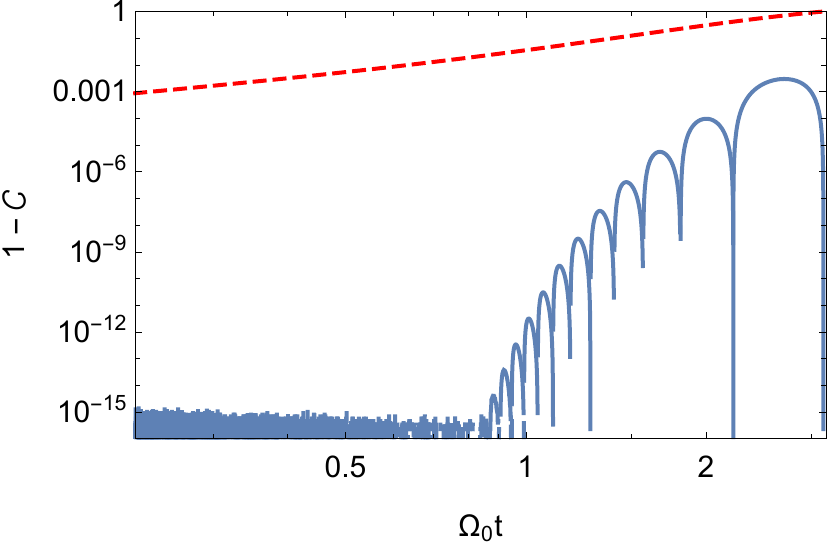}
    \caption{{Gap $1-\mathcal{C}$ between perfect coherence and that achievable with our truncated optimized states (solid blue curve) and coherent-state $\tfrac{\pi}{2}$-pulses (red dashed curve)}. Perfect coherence generation $\mathcal{C}=1$ occurs whenever $\Omega_0 t=\pi/\sqrt{n}$. At intermediate times, a few coefficients $c_n$ are sufficient for transferring near-perfect coherence; the difference between $\mathcal{C}$ and 1 reaches machine precision for $\Omega_0 t\approx \pi/\sqrt{13}$. Coherent states with $\alpha=\sqrt{\bar{n}}$, in comparison, have failure rates that only scale with $\tfrac{1}{\alpha^2}$ (i.e., with $t^2$). All scales are logarithmic.}
    \label{fig:1-coherences}
\end{figure}

The recursion relation \eqref{eq:recursion} only truncates at the aforementioned discrete set of times.
We can manually truncate the recursion relation at $n_\text{max}=\left\lceil\left(\frac{\pi}{\Omega_0 t}\right)^2\right\rceil$ to find field states that produce \textit{almost} perfect coherence due to their maximization of the Cauchy-Schwarz inequality. {While truncating at smaller or larger values of $n_\text{max}$ is still conducive to coherence generation, our choice of $n_\text{max}$ is as close as possible to a $\pi$ pulse for the highest-excitation manifold, avoiding extraneous generation of the $\ket{g}\otimes\ket{n_\text{max}}$ state.}
For times corresponding to $n_\text{max}$ as small as two, the coherence obeys $\mathcal{C}\geq 0.997$; a maximum of three or more photons in the initial field state increases the coherence beyond $\mathcal{C}\geq 0.9999$. In Fig. \ref{fig:1-coherences} we plot the deviation from unity of the coherence measure found from such a truncated series. It is readily apparent that such field states are phenomenal at transferring coherence and are much {more effective} than coherent states \add{for any finite amount of energy}. 

Another way of analysing these transcoherent states is through an expansion of $\eqref{eq:recursion}$. The peak of the distribution is found when $\left|c_{n+1}\right|=\left|c_n\right|$, which occurs precisely when 
\eq{
\sin\left(\frac{\Omega_{\bar{n}} \tilde{t}}{2}\right)=\cos\left(\frac{\Omega_{\bar{n}-1} \tilde{t}}{2}\right)
\label{eq:trigs matching}
,
} corresponding to a time defined by \eq{\Omega_0 \tilde{t}=\pi\left(\sqrt{\bar{n}+1}-\sqrt{\bar{n}}\right){\approx \frac{\pi}{2\sqrt{\bar{n}}},\quad \bar{n}\gg 1},
\label{eq:perfect time}
} which approaches a $\tfrac{\pi}{2}$ pulse in the large-$\bar{n}$ limit. 
Inverting \eqref{eq:perfect time}, the peak photon number is
\eq{
\bar{n}=\left(\frac{\pi}{2\Omega_0 \tilde{t}}-\frac{\Omega_0 \tilde{t}}{2\pi}\right)^2 \approx \frac{1}{4}\left(\frac{\pi}{\Omega_0 \tilde{t}}\right)^2,\quad \Omega_0 \tilde{t}\ll\pi.
\label{eq: optimal average n}
} When $\bar{n}$ is not an integer, it is clear that the coefficients with the largest magnitudes are $c_{\left\lfloor\bar{n}\right\rfloor}$ and $c_{\left\lceil\bar{n}\right\rceil}$. The location of this peak bears strong resemblance to the cutoff times $\Omega_0 t_n=\pi/\sqrt{n}$. Since there is only one time involved in the problem,
the field state's peak, corresponding to its average photon number, will occur at $\bar{n}\approx \tfrac{1}{4}n_\text{max}$.

Choosing a fixed $\tilde{t}$ and expanding \eqref{eq:recursion} about $\bar{n}(\tilde{t})$ uses the relation
\eq{
\frac{\cos\left[\frac{\pi}{2}\left(\sqrt{1+\bar{n}}-\sqrt{\bar{n}}\right)\sqrt{\bar{n}+\delta}\right]}{\sin\left[\frac{\pi}{2}\left(\sqrt{1+\bar{n}}-\sqrt{\bar{n}}\right)\sqrt{\bar{n}+\delta+1}\right]}\approx 1-\frac{\delta \pi}{4\bar{n}}.
}
This yields a probability distribution for the coefficients that satisfies the approximate difference equation
\eq{
\frac{c_{\bar{n}+\delta}-c_{\bar{n}}}{\delta}\approx -\frac{\delta \pi}{4\bar{n}}c_{\bar{n}} \Rightarrow \frac{\partial c(\delta)}{\partial \delta}\approx -\frac{\delta\pi}{4\bar{n}}c(\delta),
} which conforms to the number-squeezed Gaussian distribution required by \eqref{eq:number squeezed Gaussian}. Indeed, in the large-$\bar{n}$ limit, we have verified numerically that the probability distributions $\left\{\left|c_n\right|^2\right\}$ quickly converge to Gaussians with mean $\bar{n}$ and variances $\sigma^2=\tfrac{2}{\pi}\bar{n}$. We show in Appendix \ref{app:how much squeezing} that the perfect $(2k+1)\tfrac{\pi}{2}$ pulses always approach Gaussian distributions that are number squeezed by a factor of $(2k+1)\tfrac{\pi}{2}$.

In terms of average photon number, the optimal times approach 
\eq{
\Omega_0\sqrt{\bar{n}}\tilde{t}\approx\frac{\pi}{2}.
\label{eq:quantum pi/2 pulse}
} While the optimal pulses correspond to $\pi$ pulses for the maximum photon number $n_\text{max}$, the overall pulse area tends to a $\tfrac{\pi}{2}$ pulse. This is what the classical solution mimics: the optimal average intensity is a $\tfrac{\pi}{2}$ pulse because of the large-$\bar{n}$ limit of \eqref{eq: optimal average n}. The fundamental constraint for perfect coherence generation is the $\pi$ pulse of \eqref{eq:quantum pi pulses}; we recover \eqref{eq:quantum pi/2 pulse}'s $\tfrac{\pi}{2}$-pulse constraint in the classical limit.

As per Appendix \ref{app:undo pulse}, these pulses can be exactly undone by allowing for a free atomic evolution that acquires a phase $\omega t -\pi$ in \eqref{eq:ground to atomic coherent} (i.e., waiting a time $\omega \tau=\pi-\omega t+2\pi k$ for $k\in\mathds{Z}$) before turning on the JCM interaction to enact a reverse quantum $\tfrac{\pi}{2}$ pulse, or by using a quantum kick to achieve the same phase evolution \cite{Morigietal2002}. Because the atom and field are separable, the reverse pulse can even be enacted over a different period $t_{n^\prime}$ than the initial $\tfrac{\pi}{2}$ pulse.
Our idealized $\tfrac{\pi}{2}$ pulses can now be used for Ramsey interferometry, quantum information processing, and more, absent the residual atom-field entanglement generated by standard $\tfrac{\pi}{2}$ pulses.

\section{Application of transcoherent states to quantum catalysis}

One pertinent application of these ideal states is the quantum catalysis scheme proposed by Ref. \cite{Aberg2014} and recently refined by Refs. \cite{Messingeretal2020,KollasBlekos2019}. The main idea is to use the same field state more than once to create atomic coherent states. To do so, one must consider the likelihood not only of a given field state successfully creating an atomic coherent state but also of the updated field state successfully creating an atomic coherent state. In this language it is more natural to consider the probability measure \eqref{eq:probability measure} than the related coherence measure $\mathcal{C}$. Since our ideal field states are infinitely more successful at creating atomic coherent states (i.e., the probability of failure for any other state is infinitely higher than for the ideal states), we predict the former to be more robust to measurement backaction than regular coherent states and thus able to successfully catalyze more coherent atomic state generation events.

 {Strictly, catalysts experience no backaction; in truth}, a field state with finite energy can only be used a finite number of times before degrading, due to the change in field state following coherence transfer. This is the case when the atoms start in their ground states, due to conservation of energy, and is also the case when the atoms start in their excited states due to the finite total amount of coherence. We proceed by discussing coherent catalysis in the context of atomic ground states, which are easier to implement experimentally; incidentally, the field states seem to be more robust to measurement backaction when the atoms are initially in their ground states.

\begin{figure}
    \centering
    \includegraphics[width=0.5\textwidth]{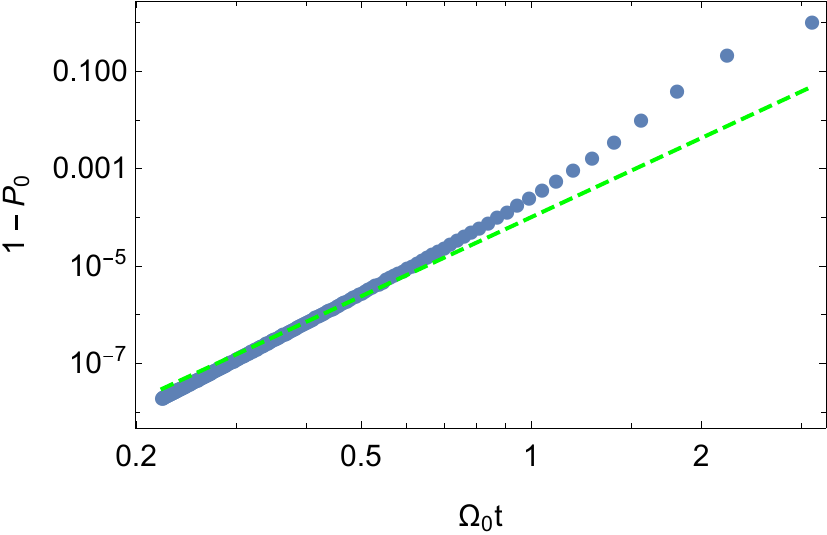}
    \caption{Usefulness of an ideal, transcoherent field state in creating a second maximally coherent atomic state. Plotted {is the failure probability $1-P_0$} of creating a second atomic coherent state after having perfectly transferred coherence to a first atom over a duration $\Omega_0 t=\pi/\sqrt{n}$, on a logarithmic scale (blue dots). The failure rates are incredibly small and decrease precipitously with decreasing time and increasing average photon number; for scale we include the power law $10^{-4}\left(\Omega_0 t\right)^{5.4}$ (dashed green line).
    }
    \label{fig:successive success ideal}
\end{figure}

After evolving to state \eqref{eq:ground to atomic coherent}, can the new field state 
\eq{
\frac{1}{\sqrt{\mathcal{N}}}\sum_n c_n \eu^{-\iu n \omega t}\cos\left(\frac{\Omega_{n-1}t}{2}\right)\ket{n}
}
be reused to create another atomic coherent state? It will never be able to do so perfectly because it does not satisfy \eqref{eq:recursion}. We evolve this state using \eqref{eq:JCM solution from ground}, calculate the probability of successfully creating a second coherent state using \eqref{eq:probability measure}, and maximize the result with respect to a new interaction time $t$. The results for these optimizations are plotted in Fig. \ref{fig:successive success ideal}. The probability of success for the second catalysis event is significantly greater than the probability that a coherent state catalyzes even a single event in the initial time $\Omega t$ (Fig. \ref{fig:1-coherences}), but the probability of failure is now nonzero at all subsequent times. To go further with coherent catalytic models, one must quantify the probability of success of a series of catalysis events each conditioned on the success of the previous event, which is no longer guaranteed.

\begin{figure}
    \centering
    \includegraphics[width=0.5\textwidth]{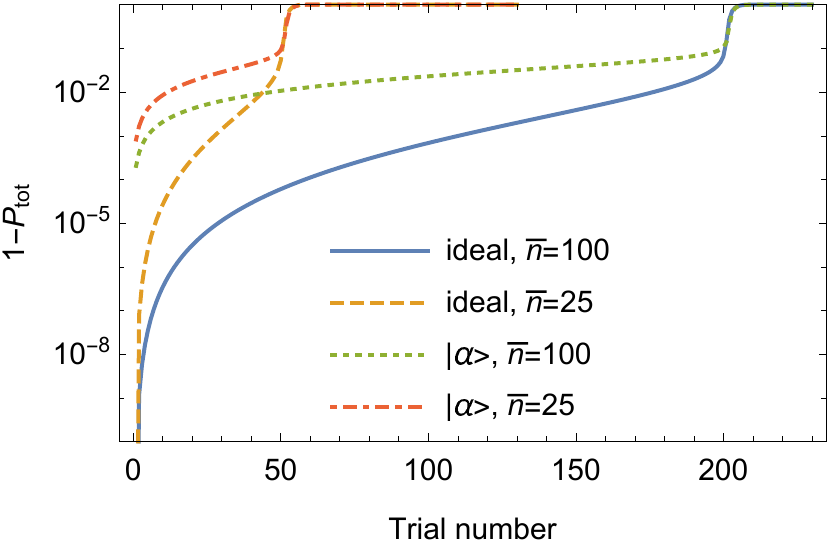}
    \caption{
    Probability that a single field state successfully ``catalyzes'' a sequence of reactions taking a series of atoms in state $\ket{g}$ to atomic coherent states. The initial field states have average photon number of either $\approx 100$, to permit comparison with the scheme from Refs. \cite{Messingeretal2020,KollasBlekos2019} using coherent states, or $\approx 25$. The ideal initial field states found using \eqref{eq:recursion} significantly outperform coherent states. The reactions maintain a $90\%$ total success probability for $2\bar{n}$ reactions.
    }
    \label{fig:total successive success n25 and n100}
\end{figure}

We iterate this process to determine the overall probability of consecutively creating $N$ atomic coherent states. Conservation of energy dictates that a field state with  a fixed initial energy $\bar{n}$ can act as a catalyst for the process
\eq{
\ket{g}\to\frac{\ket{g}+\ket{e}}{\sqrt{2}}
\label{eq:g to coh reaction}
} at most $N=2\bar{n}$ times. While both coherent states and our ideal states are useful for all $N$ reactions, the ideal states outperform coherent states for all $N$ reactions. The overall failure probability for transcoherent states is on the order of $\frac{1}{\bar{n}}$ times the failure probability for coherent states (Figs. \ref{fig:total successive success n25 and n100}). 
States that perfectly transfer coherence to atoms are extremely useful for coherent catalysis.

\add{Given that transcoherent states outperform coherent states for quantum catalysis, it is natural to ask whether transcoherent states can, in turn, be surpassed. This requires fixing a figure of merit: if the goal is to generate as much coherence as possible within a single atom for a fixed energy $\bar{n}$, transcoherent states are optimal; if the goal is to generate as much coherence as possible in $2\bar{n}$ atoms, transcoherent states outperform coherent states but may be surpassed by another set of states for some values of $\bar{n}$; and, if the goal is to optimize the amount of coherence transferred for a fixed energy in a fixed amount of time, then transcoherent states are optimal for sufficient energies and sufficient times but perhaps could be outperformed when there is insufficient energy to achieve perfect coherence in the desired time. All of these questions and their offshoots merit further study and certainly benefit from studies beyond coherent states.}

\add{The existence of transcoherent states elicits many subsequent questions. For example, how do transcoherent states compare to coherent states in generalizations to dynamics beyond the JCM, and are there other states that outperform both of the former in the generalized dynamics? One such generalization is to dynamics beyond the rotating-wave approximation on which the JCM is contingent. In that case, energy nonconservation will allow creation of both the state $\ket{e,n_\text{max}+2}$ in addition to the usual $\ket{e,n_\text{max}+1}$, which will both simultaneously require probability amplitude zero; perfect coherence transfer will therefore be unlikely. This only becomes relevant when the atom-field detuning strays from zero, which we have assumed to hold here, so an intermediate followup question is how to optimally transfer coherence within the rotating-wave approximation but with nonzero detuning. We do not expect ideal coherence transfer to be possible for nonzero detuning due to the required truncation condition for the field states, but we do expect our method to yield states that generate more coherence than canonical coherent states.

Further considerations include the effects of dissipation on the atom and decoherence on the field state. We expect neither process to preferentially affect coherent versus transcoherent states because both sets of states change during their interactions with an atom, making the initial character of the field state less relevant over time. 
One could consider the effect of decoherence on the field state \textit{prior to} its interaction with the atom: an amplitude-damping channel $\ha\to \sqrt{1-\eta^2}\ha+\eta\hb$ for vacuum mode $\hb$ and small $\eta$ acting on transcoherent states makes their failure rates increase with $\eta^2$. Conversely, coherent states effectively suffer an energy decrease $\alpha\to\sqrt{1-\eta^2}\alpha$, leading to a timing mismatch $\Omega\Delta t\approx \tfrac{\pi}{\alpha}\frac{\eta^2}{2}$. The sensitivity to timing has an error rate on the order of $\Delta t^2$ for both coherent and transcoherent states, so coherent states are only hindered by the decoherence induced by an amplitude damping channel to order $\eta^4$. This confirms the unsurprising result that coherent states are more stable to preparation errors than transcoherent states; still, amplitude damping of transcoherent states degrades their squeezing toward coherent states, so any finite errors in the preparation still yield states that are partially number squeezed and can therefore outperform coherent states. Further studies are warranted to investigate optimal stability to errors occurring during the atom-field interaction and to generalize this interaction to traveling fields \cite{HolmMolmer2020}.}

{
In conclusion, it is possible to overcome the limitations of canonical coherent states for generating atomic coherence, by using tailored states of light we term ``\textit{trans}coherent states'' because they surpass coherent states in their ability to transfer coherence.
These states generate atomic superpositions} with maximal dipole moments and without residual atom-field entanglement, fulfilling the broken promise of semiclassical $\tfrac{\pi}{2}$ pulses in a fully-quantized domain.

The idealized pulses can create atomic superpositions, useful for quantum tasks in information processing, thermodynamics, and beyond, in arbitrarily short times.
They can be reused $\mathcal{O}(\bar{n})$ times with excellent success rates, extending the notion of quantum catalysis \cite{Aberg2014,Messingeretal2020}, and
they can replace standard $\tfrac{\pi}{2}$ pulses, with applications including higher-precision measurements in Ramsey interferometry \cite{Ramsey1950} and its many applications \cite{Bruneetal1996,Dowling1998,Vuthaetal2017,Breweretal2019}.
All of the other tasks requiring coherent atomic states can similarly benefit from our schemes for maximal coherence transfer. 

The ideal states are well-suited to experimental generation. They are well-approximated in the large-$\bar{n}$ limit by coherent states that are number squeezed by $-10\log_{10}\tfrac{2}{\pi}\approx 2$ dB, which is readily achievable \add{with quadrature squeezing} \cite{Wuetal1986,Vahlbruchetal2016}\add{. This is because $2$ dB of number squeezing resembles quadrature squeezing for states with an average of more than a few photons, which can be seen by the negativity of a transcoherent state's Wigner function, and thereby its nonGaussianity, shrinking exponentially with the average number of photons in the state}. They can also be created by tailoring a few field excitations in the small-$\bar{n}$ limit 
using circuit QED \cite{Raimondetal2001}.
For quantized $(2k+1)\tfrac{\pi}{2}$ pulses the squeezing requirement increases by $10\log_{10}(2k+1)$ dB.
While answering the fundamental question of what limits coherence transfer in the JCM, these results open the door to numerous exciting applications\rem{ that are poised to revolutionize the modern world}. We hope to see these intriguing states implemented experimentally in the near future.

\begin{acknowledgments}
{We acknowledge a question Zlatko Minev posed at the opening reception of the 2019 Rochester Conference on Coherence and Quantum Optics, and subsequent discussions with him and Howard Wiseman, which set us thinking about this problem; and Daniel James for pointing us to relevant literature.  This work was supported by NSERC Discovery grants.
AZG additionally acknowledges funding from an NSERC
Alexander Graham Bell Scholarship, the Walter C.
Sumner Foundation, and the Lachlan Gilchrist Fellowship Fund.  AMS is a Fellow of CIFAR.}
\end{acknowledgments}

\onecolumngrid
\appendix 

\section{Idealized $(2k+1)\tfrac{\pi}{2}$ pulses are number squeezed by $(2k+1)\tfrac{\pi}{2}$}
\label{app:how much squeezing}
We show that the idealized pulses that perfectly generate coherent atomic states approached number-squeezed Gaussians, where the amount of number squeezing is exactly equal to the pulse area.
\subsection{Atom initially in $\ket{g}$}
The recursion relation \eqref{eq:recursion} is stationary when $\sin\left(\frac{\Omega_n \tilde{t}}{2}\right)=\cos\left(\frac{\Omega_{n-1} \tilde{t}}{2}\right)$. This happens when
\eq{
\Omega_0 \tilde{t}&=\left(4k+1\right)\pi\left(\sqrt{n+1}-\sqrt{n}\right),\quad k\in\mathbb{Z}\\
&\approx \left(2 \pi  k+\frac{\pi }{2}\right) \sqrt{\frac{1}{n}},\quad n\gg 1,
} and this maximum happens at $\bar{n}$ when the lowest-excitation manifold undergoes a $2k\pi$ pulse and the highest a $(2k+1)\pi$ pulse [viz., a $(4k+1)\tfrac{\pi}{2}$ pulse on average].
Expanding the recursion relation \eqref{eq:recursion} around $\bar{n}$ yields
\eq{
\frac{\cos\left[\frac{\Omega_0\tilde{t}}{2}\sqrt{\bar{n}+\delta}\right]}{\sin\left[\frac{\Omega_0\tilde{t}}{2}\sqrt{\bar{n}+\delta+1}\right]}
\approx 1-\frac{  \delta \pi (4 k+1)}{4 \bar{n}}.
}
The coefficients satisfy the approximate difference equation
\eq{
\frac{c_{\bar{n}+\delta}-c_{\bar{n}}}{\delta}\approx -\frac{\delta \pi(4 k+1)}{4\bar{n}}c_{\bar{n}} \quad \Rightarrow \quad \frac{\partial c(\delta)}{\partial \delta}\approx -\frac{\delta\pi(4 k+1)}{4\bar{n}}c(\delta) \quad\Rightarrow\quad \left|c(\delta)\right|^2\propto \exp\left[-\frac{(4k+1)\pi}{4\bar{n}}\delta^2\right];
} i.e., the probability distribution is Gaussian with $\sigma^2=\tfrac{2\bar{n}}{(4k+1)\pi}$. These $(4k+1)\tfrac{\pi}{2}$ pulses are therefore number squeezed by $(4k+1)\tfrac{\pi}{2}$. 

\subsection{Atom initially in $\ket{e}$}
The recursion relation \eqref{eq:recursion starting e} is stationary when $\sin\left(\frac{\Omega_{n-1} \tilde{t}}{2}\right)=-\cos\left(\frac{\Omega_{n} \tilde{t}}{2}\right)$. This happens when
\eq{
\Omega_0 \tilde{t}&=\left(4k+3\right)\pi\left(\sqrt{n+1}-\sqrt{n}\right),\quad k\in\mathbb{Z}\\
&\approx \left(2 \pi  k+\frac{3\pi }{2}\right) \sqrt{\frac{1}{n}},\quad n\gg 1,
} and this maximum happens at $\bar{n}$ when the lowest-excitation manifold undergoes a $(2k+1)\pi$ pulse and the highest a $(2k+2)\pi$ pulse [viz., a $(4k+3)\tfrac{\pi}{2}$ pulse on average].
Expanding the recursion relation \eqref{eq:recursion starting e} around $\bar{n}$ yields
\eq{
\frac{-\sin\left[\frac{\Omega_0\tilde{t}}{2}\sqrt{\bar{n}+\delta}\right]}{\cos\left[\frac{\Omega_0\tilde{t}}{2}\sqrt{\bar{n}+\delta+1}\right]}
\approx 1-\frac{  \delta \pi (4 k+3)}{4 \bar{n}}.
}
The coefficients satisfy the approximate difference equation
\eq{
\frac{c_{\bar{n}+\delta}-c_{\bar{n}}}{\delta}\approx -\frac{\delta \pi(4 k+3)}{4\bar{n}}c_{\bar{n}} \quad \Rightarrow \quad \frac{\partial c(\delta)}{\partial \delta}\approx -\frac{\delta\pi(4 k+3)}{4\bar{n}}c(\delta) \quad\Rightarrow\quad \left|c(\delta)\right|^2\propto \exp\left[-\frac{(4k+3)\pi}{4\bar{n}}\delta^2\right];
} i.e., the probability distribution is Gaussian with $\sigma^2=\tfrac{2\bar{n}}{(4k+3)\pi}$. These $(4k+3)\tfrac{\pi}{2}$ pulses are therefore number squeezed by $(4k+3)\tfrac{\pi}{2}$. 

\section{Quantum $\tfrac{\pi}{2}$ pulses can be reversed}
\label{app:undo pulse}
The quantum pulses that perfectly catalyze the reaction ground-to-coherent-atomic-state reaction \eqref{eq:g to coh reaction} can be perfectly reversed. After the first pulse, the atom-field state is separable \eqref{eq:recursion}-\eqref{eq:ground to atomic coherent}: 
\eq{
\ket{\psi(t)}=\frac{\ket{g}+\ket{e}}{\sqrt{2}}\otimes \frac{1}{\sqrt{\mathcal{N}}}\sum_n c_n \eu^{-\iu n \omega t} \cos\left(\frac{\Omega_{n-1} t}{2}\right)\ket{n},\quad
c_{n+1}=c_n\frac{\iu \eu^{ \iu \omega t}\cos\left(\frac{\Omega_0 t}{2}\sqrt{n}\right)}{\sin\left(\frac{\Omega_0 t}{2}\sqrt{n+1}\right)}.
} Allowing the atom to freely evolve for a time $\tau$, inspired by Ramsey interferometry, yields 
\eq{
\ket{\psi(t;\tau)}=\frac{\ket{g}+\eu^{-\iu \omega \tau}\ket{e}}{\sqrt{2}}\otimes \frac{1}{\sqrt{\mathcal{N}}}\sum_n c_n \eu^{-\iu n \omega t} \cos\left(\frac{\Omega_{n-1} t}{2}\right)\ket{n}.
} Physically, this requires providing a new field state that has not evolved for a time $\tau$ to comprise the reverse $\tfrac{\pi}{2}$ pulse. The new field state does not have to correspond to the same $t_n$ as the forward $\tfrac{\pi}{2}$ pulse and so can have a different average energy than the first pulse.

Turning the interaction on for another time $T$ leads to
\eq{
\ket{\psi(t;\tau;T)}&\propto 
c_0\ket{g}\otimes\ket{0}+\sum_{n=0}^{\infty}c_{n+1}\eu^{-\iu\omega\left[(n+1) t+  n T\right]}\cos\left(\frac{\Omega_n t}{2}\right)\\
&\hspace{6cm}\times
\left[	
\cos\left(\frac{\Omega_n T}{2}\right)\ket{g}\otimes\ket{n+1}-\iu \sin\left(\frac{\Omega_n T}{2}\right)\ket{e}\otimes\ket{n}
\right]\\
&\quad +
c_n \eu^{-\iu\omega\left[\tau+ n t+ (n+1) T\right]}\cos\left(\frac{\Omega_{n-1}t}{2}\right)
\left[-\iu\sin\left(\frac{\Omega_n T}{2}\right)\ket{g}\otimes\ket{n+1}+\cos\left(\frac{\Omega_n T}{2}\right)\ket{e}\otimes\ket{n}\right].
}
The overlap of this state with any excited state $\ket{e}\otimes\ket{n}$ is proportional to
\eq{
& c_{n+1}\eu^{-\iu \omega t}\cos\left(\frac{\Omega_n t}{2}\right)
\left[-\iu \sin\left(\frac{\Omega_n T}{2}\right)
\right] +
\eu^{-\iu\omega\tau-\iu \omega T}c_n \cos\left(\frac{\Omega_{n-1}t}{2}\right)
\left[\cos\left(\frac{\Omega_n T}{2}\right)\right]
\\
&=c_n\cos\left(\frac{\Omega_{n-1}t}{2}\right)\left[\frac{1}{\sin\left(\frac{\Omega_n t}{2}\right)}\cos\left(\frac{\Omega_n t}{2}\right)
\sin\left(\frac{\Omega_n T}{2}\right)
+
\eu^{-\iu\omega\tau-\iu\omega T}
\cos\left(\frac{\Omega_n T}{2}\right)
\right].
}
This clearly vanishes whenever the second pulse has the same duration as the first $T=t$ and the free evolution time interferes perfectly with the atomic transition frequency $\omega(\tau+t)=\pi +2\pi k,\, k\in \mathds{Z}$.

That the reverse pulse is viable after large $\tau$ enables enhanced Ramsey interferometry with entanglement-free pulses.
Since the states are initially separable, any reverse pulse with $\Omega_n t=\pi$ can be used for the backward evolution.

\twocolumngrid

%

\end{document}